# OPTICALLY-DETECTED LONG-LIVED SPIN COHERENCE IN MULTILAYER SYSTEMS: DOUBLE AND TRIPLE QUANTUM WELLS

S. Ullah[1][*], G. M. Gusev[1], A. K. Bakarov[2], F. G. G. Hernandez[1]

[1] Instituto de Física, Universidade de São Paulo, Caixa Postal 66318 - CEP 05315-970, São Paulo, SP, Brazil

[2] Institute of Semiconductor Physics and Novosibirsk State University, Novosibirsk 630090, Russia

[*]To whom Correspondence should be addressed: saeedullah.phy@gmail.com

## ABSTRACT

In this contribution, we investigated the spin coherence of high-mobility dense two-dimensional electron gases confined in multilayer systems. The dynamics of optically-induced spin polarization was experimentally studied employing the time-resolved Kerr rotation and resonant spin amplification techniques. For both the double and triple quantum wells doped beyond the metal-insulator transition, where the spin coherence is greatly suppressed, we found remarkably long spin lifetimes limited by the Dyakonov-Perel mechanism and spin hopping process between the donor sites as well as the spread of ensemble $g$-factor. The double quantum well structure yields a spin lifetime of 6.25 ns at $T$ = 5 K while the triple quantum well shows a spin lifetime exceeding 25 ns at $T$ = 8 K.

## 1. INTRODUCTION

The device concepts in semiconductor nanostructures rely mainly on the efficient generation of spin polarization, its manipulation, and detection [1]. The fabrication of these future devices could benefit from the long-lived spin coherence [2-4]. A number of experimental techniques have been developed to study the spin polarization dynamics and spin relaxation mechanisms in semiconductor nanostructures. Among those techniques, the double-pulsed pump-probe technique is one of the most reliable tools [5-10]. The principle of this technique is as follow: A circularly-polarized light of laser pulse (the pump) incident normal to the sample structure creates the spin-polarized electrons with the spin vector along the sample growth direction and a relatively weak linearly-polarized probe pulse, from the same laser, is used to detect the spin polarization dynamics of electrons in two-dimensional electron gas (2DEG). Using this technique, the spin dynamics can be studied in time intervals shorter than the repetition period of the pump pulses, which is usually ~13 ns for commonly used Ti:sapphire laser.

It has been reported in the bulk semiconductors [7], quantum dots (QDs) [11], and quantum wells (QWs) [2,3,5] that the spin lifetime ($T_2^*$) exceeds the laser repetition and attains several nanoseconds in the transverse magnetic field. When $T_2^*$ becomes equal to or greater than the laser repetition period, then the direct determination by time-resolved methods becomes inapplicable. In such situation, the procedure of resonant spin amplification (RSA) [7] can be used for the determination of $T_2^*$. For the RSA measurements, the pump-probe delay ($\Delta t$) is kept fixed while the dependence of Kerr rotation (KR) signal on experimental parameters can be studied by sweeping the external magnetic field in the milli-Tesla range. When the laser repetition time becomes multiple to the period of spin precession a series of sharp resonance peaks as a function of magnetic field can be observed. The spacing of those resonance peaks allows to calculate the carrier $g$-factor while their line-width points the spin lifetime.

The aim of the present work is to study the long-lived spin coherence in GaAs/AlGaAs double (DQW) and triple quantum wells (TQW). These structures were selected for present investigation because such multiple quantum well systems result in the discovery of fascinating phenomena, such as, current-induced spin polarization [12], spin Hall effect [13], large anisotropic spin relaxation [14], and macroscopic transport and transverse drift of long current-induced spin coherence [15,16]. We observed a remarkably long $T_2^*$ exceeding 25 ns in the TQW. The obtained values are among the longest reported $T_2^*$ in the structures with similar doping level [17,18] and comparable to the one reported in undoped GaAs QWs [19] and low-density two-dimensional electron gases confined in CdTe QWs [5].

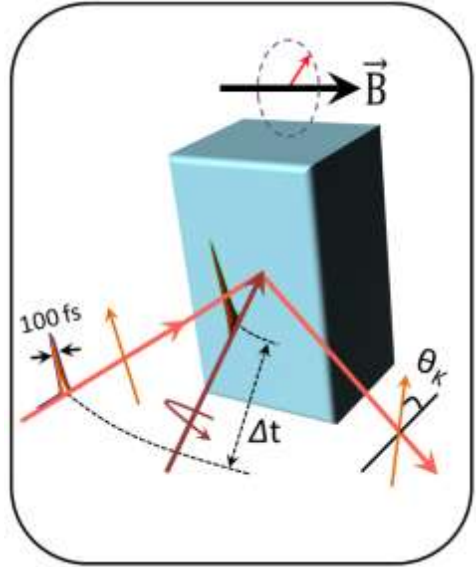

**Fig. 1** Scheme of the time-resolved pump-probe technique: The spin polarization was induced by a circularly polarized pump and was monitored by a relatively weak linearly polarized probe pulse.

The manuscript is organized as follows. Section 2 is devoted to the materials and experiment. Section 3 presents the experimental results of spin dynamics reported in two different samples while concluding remarks are briefly discussed in section 4.

## 2. MATERIALS AND EXPERIMENT

The structures used in this study are GaAs/AlGaAs double and triple quantum wells grown by molecular-beam epitaxy (MBE) on a (001)-oriented GaAs substrate. Both samples are symmetrically $\delta$-doped beyond the metal-insulator transition (5 $\times$ 10$^{10}$ cm$^{-2}$ for GaAs QWs [17]). The DQW structure is a 45-nm wide GaAs well with total electron sheet density $n_s = 9.2 \times 10^{11}$ cm$^{-2}$ and low-temperature mobility $\mu = 1.9 \times 10^6$ cm$^2$/Vs. Due to the large well width and high electron density, the Coulomb repulsion of electrons results in a DQW configuration with two occupied subbands with subband separation of $\Delta_{12} = 1.4$ meV [20]. The TQW structure has a 22-nm thick central well and two 10-nm thick side wells separated by 2-nm thick AlGaAs barriers. The tunneling of electron states, due to thin barriers, results in three populated subbands with separation energies $\Delta_{12} = 1.0$ meV, $\Delta_{23} = 2.4$ meV, and $\Delta_{13} = 3.4$ meV [21]. To keep the central well populated its width was kept wider than lateral wells because the electron density mostly concentrates in the side wells due to electron repulsion and confinement. The estimated density in lateral wells is 35 % larger than in the central wells. The calculated band structure and charge density of occupied subbands for both the DQW and TQW are demonstrated in Ref. [2].

We used time-resolved Kerr rotation (TRKR) [22] and resonant spin amplification [7] to study the coherent spin dynamics in multilayer structures. For the optical excitation, we used a Ti-sapphire laser with 100 fs pulse duration and a repetition period of $t_{rep} = 13$ ns. The samples were kept in a He flow cryostat and exposed to an external magnetic field applied normal to the light beam direction (Voigt geometry) as shown in Fig. 1. The spin polarization along the quantum well growth direction was induced by circularly polarized pulses focused onto a spot of ~50 $\mu$m. For most of the experiments, except power dependence, we used a pump power of 1 mW which gives rise to a photogenerated density of $2 \times 10^{11}$ cm$^{-2}$. The time evolution of optically generated spins was studied through the rotation of linearly polarized probe reflected by the sample and detected with a balanced bridge.

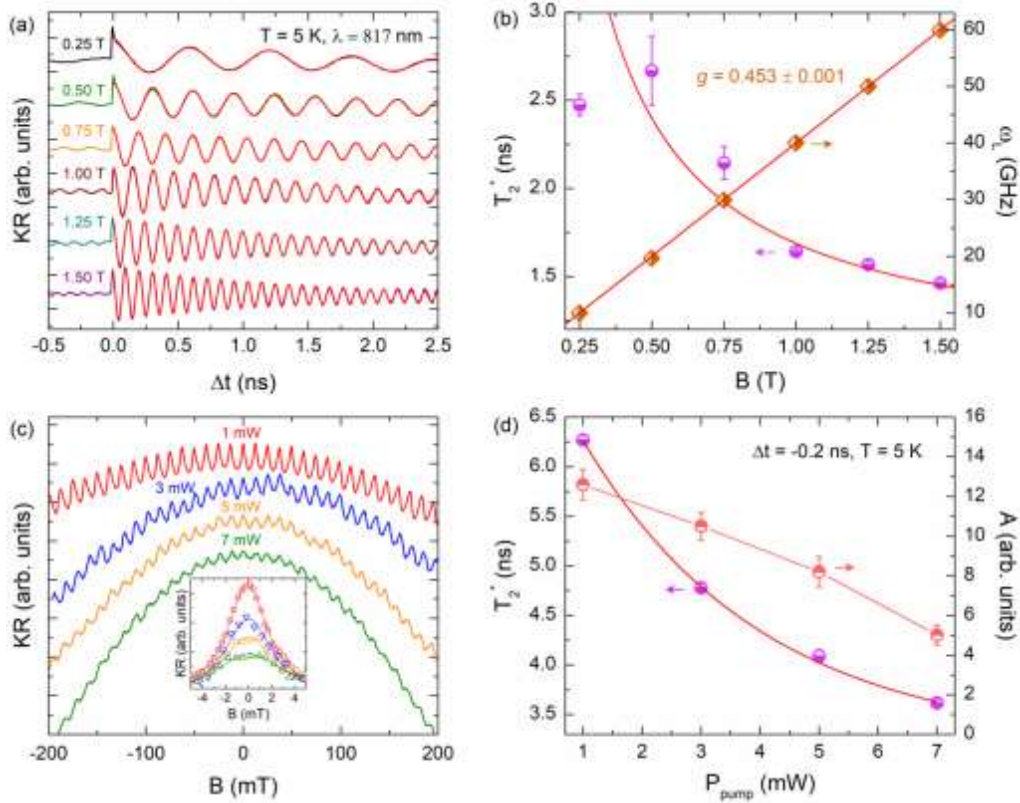

**Fig. 2** (a) TRKR traces fitted to Eq. 1 at various magnetic fields measured at $T = 5$ K for the DQW. (b) Spin precession frequency and lifetime versus the applied magnetic field. (c) RSA for the DQW as a function of optical pump power. (d) The corresponding $T_2^*$ and amplitude of the zero-field peak. The inset shows the Lorentzian fit of the zero field peaks.

## 3. RESULTS AND DISCUSSION

Fig. 2 (a) shows a series of TRKR curves recorded at different magnetic fields ranging from 0.25 T to 1.5 T. The experimental conditions were selected for maximum Kerr signal. The observed oscillations are associated with the precession of excited electron spins about an applied external magnetic field. To retrieve information about the spin coherence time and precession frequency ($\omega_L = g\mu_B B/\hbar$) the curves were fitted to an exponentially damped harmonic function:

$$\Theta_K = A \exp(-\Delta t/T_2^*) \cos(g\mu_B B \Delta t/\hbar + \varphi) \qquad (1)$$

where $A$ is the initial spin polarization induced by the pump, $g$ is the electron $g$-factor, $\mu_B$ is the Bohr magneton, $\hbar$ is the reduced Planck constant, $B$ is the applied magnetic field and $\varphi$ is the oscillation phase. The obtained $T_2^*$ (half-filled circles) and $\omega_L$ (half-filled diamonds) as a function of the applied magnetic field are shown in Fig. 2 (b).

As expected $\omega_L$ versus $B$ follows a linear dependence which is typical for electrons [2], however, for holes, the band mixing may result in non-linearities as reported for $In_xGa_{1-x}As/GaAs$ QWs [23]. The linear interpolation of data resulted in a $g$-factor (absolute value) of $0.453 \pm 0.001$ which is close to the reported $g$ value for bulk GaAs. The increase of magnetic field results in a strong $T_2^*$ reduction which is well described by $1/B$-like dependence [7,24]. For increasing field, the spin dephasing was dominated by the Dyakonov-Perel (DP) mechanism [25] as well as the variations in ensemble $g$-factor. For a spread of ensemble $g$-factor $\Delta g$, the spin lifetime is given by $T_2^* = \sqrt{2}\hbar/\Delta g\mu_B B$ which allows to estimate the size of inhomogeneity by fitting with the data. Such a fit to the data, shown by a solid red curve in Fig. 2 (b), gives $\Delta g = 0.0026$.

In the magnetic field dependence of TRKR, we observed that the oscillations on positive delay were accompanied by oscillations at negative delay also which reflect the long-lived spin coherence persisting between successive pulses. Both (the positive and negative delay oscillations) have the same origin since the negative ones were positive of the previous pulse, and

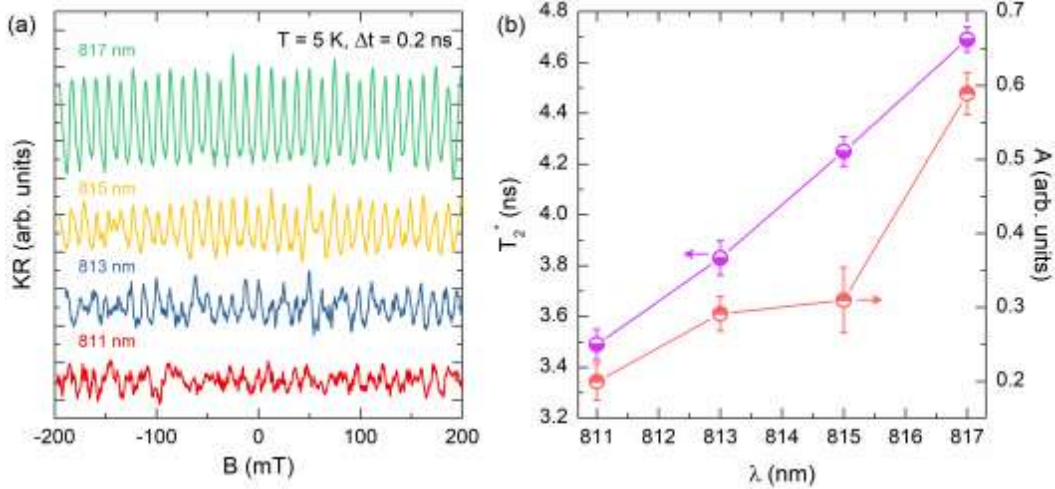

**Fig. 3** (a) RSA measurements for various pump-probe wavelengths recorded at $T = 5$ K and $\Delta t = 0.2$ ns. (b) Spin lifetime and amplitude of zero-field peaks extracted from (a).

both share the same decay time. Thus, to get a realistic $T_2^*$ we referred to the RSA technique at a short negative delay (which is the longest possible positive delay). Fig. 2 (c) shows the RSA curves for the DQW recorded at various pump power by sweeping the magnetic field in a range of -200 mT to +200 mT while keeping $\Delta t = -0.2$ ns fixed. One can clearly see that the amplitude of RSA peaks decreases and getting broader with increasing field due to the $g$-factor variation in the measured ensemble. The half-width ($B_{1/2}$) of those peaks obeying periodic condition $\Delta B = hf_{rep}/g\mu_B$, allow us to retrieve $T_2^*$ by using the Lorentzian (Hanle) model [7]:

$$\Theta_K = A/[1 + (\omega_L T_2^*)^2] \qquad (2)$$

where $T_2^* = \hbar/g\mu_B B_{1/2}$. Lorentzian fit (solid lines) to the peaks centered at zero magnetic fields are displayed in the inset Fig. 2 (c). The obtained values for amplitude and $T_2^*$ are depicted in Fig. 2 (d) as a function of excitation power. With the increase of excitation power, both quantities decrease, however, $T_2^*$ display an exponential decay (solid curve). The observed $T_2^*$ reduction may be possibly due to the heating effect caused by optical excitation [2].

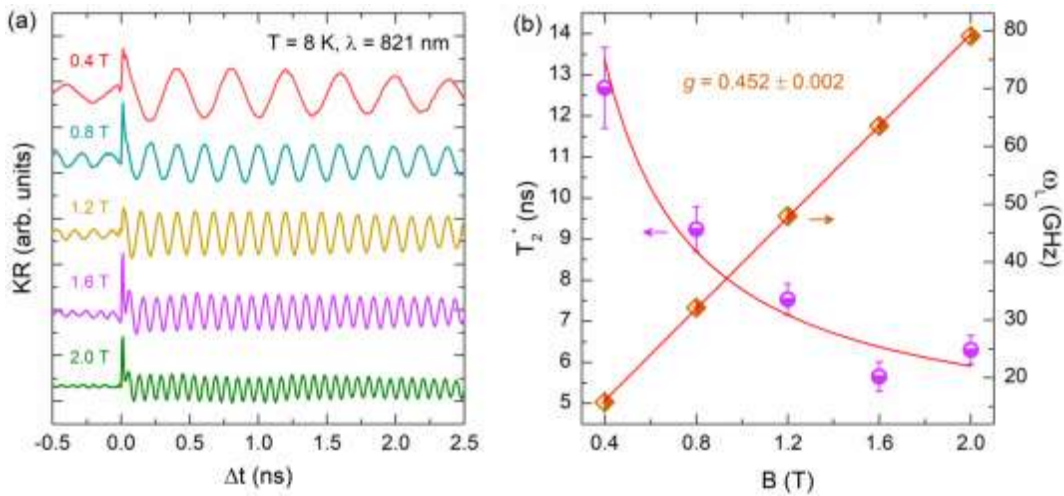

**Fig. 4** (a) TRKR curves of the TQW as a function of external magnetic field measured at $T = 8$ K. (b) Spin precession frequency and spin lifetime as a function of applied magnetic field.

Spectral dependence of Kerr signals, with the pump-probe wavelength ranging from 811 nm to 817 nm, recorded at $\Delta t = 0.2$ ns for the DQW structure are shown in Fig. 3 (a). The recorded RSA pattern, having all the peaks of comparable height, corresponds to the regime of isotropic spin relaxation. However, in the anisotropic spin relaxation, the spin components of carrier's oriented in- and out-plane relax at different rates and hence affect the amplitude of zero-th field RSA peak. The retrieved $T_2^*$ and amplitude, by fitting the data to Hanle model, increases with the excitation wavelength as plotted in Fig. 3 (b). Changing the excitation energy about 3 meV (~$2\Delta_{12}$), by increasing the pump-probe wavelength from 815 nm to 817 nm, results in a $T_2^*$ increase of less than 10 %. The observed small difference is attributed to the relatively similar charge density distribution of electrons for both subbands.

We now turn to the spin dynamics in TQW. Fig. 4 (a) shows a series of TRKR curves recorded at $T = 8$ K and $P_{pump} = 1$ mW for the various magnetic field in a range of 0.4 T to 2.0 T. The pump-probe energy was tuned to the exciton bound to the neutral donor transition [14] for maximum Kerr signal. As evidenced by negative delay oscillations, the long-lived spin coherence was observed on a time scale longer than the laser repetition period. To extract $T_2^*$ and $\omega_L$, the oscillations at positive $\Delta t$ were fitted to Eq. (1). The fitting results are displayed in Fig. 4 (b). The linear increase of spin precession frequency on increasing magnetic field indicates that the observed g value is constant in the measured range of external magnetic field (0.4-2 T). From the slope of linear fit (solid red line), we evaluated $g = 0.452 \pm 0.002$. We obtained $T_2^* = 12.7$ ns at 0.4 T which decreases with further increase of magnetic field. At low temperature, the observed $T_2^*$ reduction was attributed to the spin hopping process between the donor sites as well as the contribution of g-factor inhomogeneity. $1/T_2^*$ as a function of applied field (not shown here) follows a linear increase which is a well-known indication of the ensemble spread of g-factor originating from the inhomogeneous spin relaxation rates. From the slope, expressed by $\Delta g \mu_B/\sqrt{2}\hbar$, we obtained $\Delta g = 4.9 \times 10^{-4}$ (0.1 % of observed g-value) which is very small compared to the reported value in quantum dots. The observed small variations highlight the high structural uniformity of studied sample.

Taking into account the negative delay oscillation, in analogy to the DQW, we carried out the RSA measurement to evaluate the spin lifetime. The magnetic field was scanned in a range from -200 mT to +200 mT while adjusting $\Delta t$ such that the probe pulse arrives 240 ps before the pump pulse. Fig. 5 (a) shows multiple RSA peaks recorded for different pump-power. The observed RSA spectrum differs from that of the DQW i.e. the resonance peaks centered at zero magnetic fields are smaller in amplitude than those at the finite field. The depression of those zero-field peaks reveals the existence of spin relaxation anisotropy caused by the presence of an internal magnetic field. The spin lifetime and amplitude, extracted from the

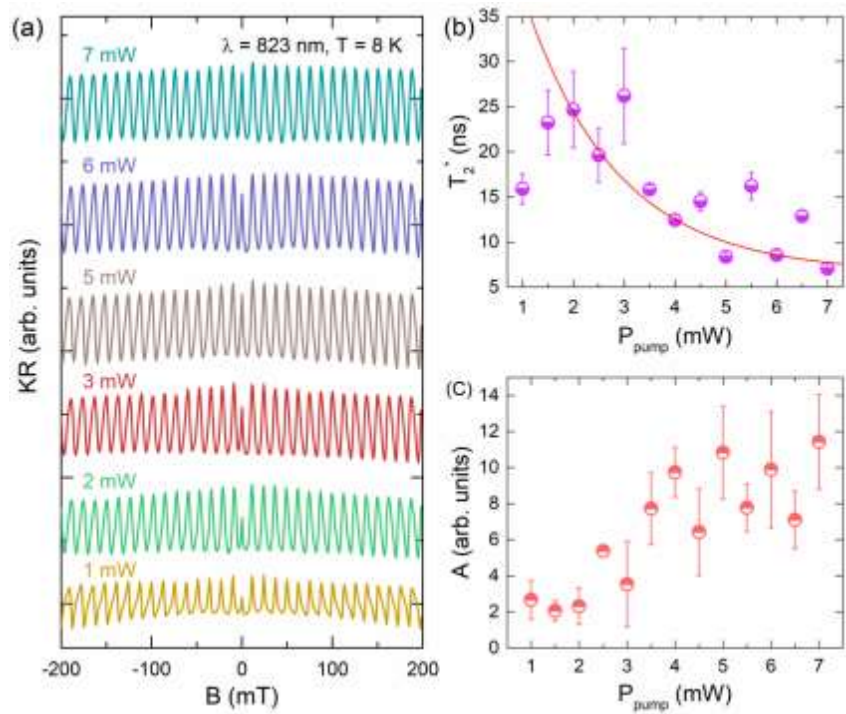

**Fig. 5** (a) RSA pattern of the TQW sample measured for various pump power at $T = 8$ K. (b) Spin lifetime and (c) amplitude extracted from (a).

Lorentzian fit to the zero-field peak, are plotted in Fig. 5 (b) and (c) as a function of pump power. We observed a remarkably long $T_2^*$ exceeding 25 ns. As evidence from RSA spectrum, the amplitude of zero-field peaks increases with high pump power and become equal to that of finite field peaks 7 mW. In contrast, $T_2^*$ decreases with pump power yielding an exponential decay. The reduction of $T_2^*$ with increasing pump power was mainly due to the heating effect caused by optical excitation [2] as well as the increased efficiency of Bir-Aronov-Pikus mechanism induced by a high density of photo-created carriers [26]. The obtained results are amongst the longest reported $T_2^*$ in samples of similar doping levels [17,18] and comparable to that of undoped GaAs QWs [19].

Fig. 6 (a) presents the energy dependence of RSA signals, with excitation wavelength ranging from 816 nm to 823 nm, measured at $\Delta t$ = - 0.24 ns for the TQW structure. Fig. 6 (b) shows the comparison between normalized zero-field peaks fitted to Lorentzian model for several wavelengths. $T_2^*$ and amplitude obtained from (b) are plotted in Fig. 6 (c). Changing the pump-probe wavelength from 816 nm to 819 nm has no effect on the spin lifetime and amplitude of zero-field peak. Increasing the wavelength beyond 820 nm, which is the origin of spin relaxation anisotropy, results in the rapid variation of amplitude and $T_2^*$. Here, the same energy variation (3 meV $\cong \Delta_{13}$), by changing wavelength from 821 to 823 nm, results in a strong $T_2^*$ reduction by almost 35 %. This large variation may be possibly due to the different charge density distribution in the first and third occupied subband.

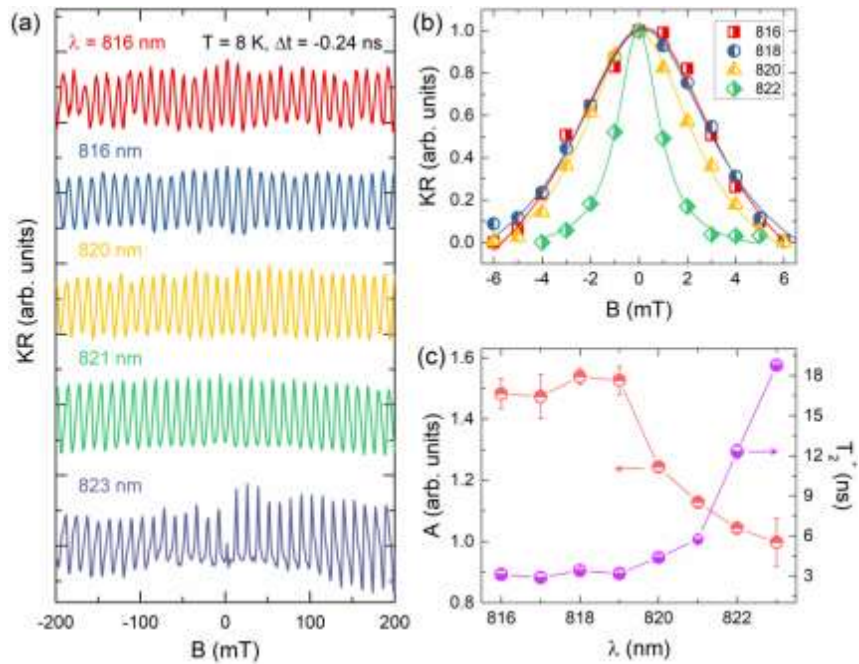

**Fig. 6** (a) Magnetic field scan of KR signal recorded for various excitation wavelengths at $T$ = 5 K and $\Delta t$ = -0.24 ns. (b) Comparison of normalized zero-field peaks fitted to Lorentzian model (c) Spin dephasing time and amplitude of zero-field peaks as a function of excitation wavelength.
.

## 4. CONCLUSIONS

In conclusion, we carried out a detailed investigation of spin relaxation in two-dimensional electron gases confined multilayer quantum wells by employing the TRKR and RSA techniques. The dependence of spin lifetime was studied as a function of experimental parameters like a magnetic field, optical pump power, and excitation wavelengths. In the TQW sample, $T_2^*$ extends to a very long time at $T$ = 8 K while the DQW structure yields a relatively small $T_2^*$. We believe that achieving the long-lived spin coherence and the wavefunction engineering in multilayer structure will open a practical path for spintronics devices.

**ACKNOWLEDGMENT**

F.G.G.H. acknowledges the financial support from Grant No. 2009/15007-5, 2013/03450-7, 2014/25981-7 and 2015/16191-5 of the São Paulo Research Foundation (FAPESP). S.U acknowledges TWAS/CNPq for financial support.